\documentclass{aa}  

\usepackage{graphicx}
\usepackage[varg]{txfonts}
\begin{document}
\title{Smoke in the Pipe Nebula: dust emission and grain growth in the starless core FeSt~1-457}
\titlerunning{Smoke in the Pipe Nebula}


   \author{Jan Forbrich
          \inst{1,2}
          \and
	  Charles J. Lada
	  \inst{2}
	  \and
	  Marco Lombardi
	  \inst{3}
	  \and
	  Carlos Rom\'an-Z\'u\~niga\inst{4}
	  \and
	  Jo\~ao Alves
	  \inst{1}
          }

   \institute{University of Vienna, Department of Astrophysics, T\"urkenschanzstra{\ss}e 17, 1180 Vienna, Austria\\
              \email{jan.forbrich@univie.ac.at}
         \and
	 Harvard-Smithsonian Center for Astrophysics, 60 Garden Street, Cambridge, MA 02138, USA
	 \and
	 University of Milan, Department of Physics, via Celoria 16, I-20133 Milan, Italy
         \and
	 Universidad Nacional Aut\'onoma de M\'exico, Instituto de Astronom\'ia, Ensenada BC 22860, Mexico
	     }



\abstract
 {The availability of submillimeter dust emission data in an unprecedented number of bands provides us with new opportunities to investigate the properties of interstellar dust in nearby clouds.} 
 {The nearby Pipe Nebula is an ideal laboratory to study starless cores. We here aim to characterize the dust properties of the FeSt 1-457 core, as well as the relation between the dust and the dense gas, using \textit{Herschel}, \textit{Planck}, 2MASS, ESO Very Large Telescope, APEX-\textit{Laboca}, and IRAM~30m data.}
 {We derive maps of submillimeter dust optical depth and effective dust temperature from \textit{Herschel} data that were calibrated against {\it Planck}. After calibration, we then fit a modified blackbody to the long-wavelength {\it Herschel} data, using the {\it Planck}-derived dust opacity spectral index $\beta$, derived on scales of 30$'$ (or $\sim$1~pc). We use this model to make predictions of the submillimeter flux density at 850~$\mu$m, and we compare these in turn with APEX-{\it Laboca} observations. Our method takes into account any additive zeropoint offsets between the {\it Herschel}/{\it Planck} and {\it Laboca} datasets. Additionally, we compare the dust emission with near-infrared extinction data, and we study the correlation of high-density--tracing N$_2$H$^+$ emission with the coldest and densest dust in FeSt 1-457.}
 {A comparison of the submillimeter dust optical depth and near-infrared extinction data reveals evidence for an increased submillimeter dust opacity at high column densities, interpreted as an indication of grain growth in the inner parts of the core. Additionally, a comparison of the {\it Herschel} dust model and the {\it Laboca} data reveals that the frequency dependence of the submillimeter opacity, described by the spectral index $\beta$, does not change. A single $\beta$ that is only slightly different from the {\it Planck}-derived value is sufficient to describe the data, $\beta=1.53\pm0.07$. We apply a similar analysis to Barnard\,68, a core with significantly lower column densities than FeSt 1-457, and we do not find evidence for grain growth but also a single $\beta$. Finally, our previously reported finding of a correlation of N$_2$H$^+$ emission with lower effective dust temperatures is confirmed for FeSt 1-457 in mapping observations.}
 {While we find evidence for grain growth from the dust opacity in FeSt 1-457, we find no evidence for significant variations in the dust opacity spectral index $\beta$ on scales $0.02<x<1$~pc (or $36''<x<30'$). The correction to the {\it Planck}-derived dust $\beta$ that we find in both cases is on the order of the measurement error, not including any systematic errors, and it would thus be reasonable to directly apply the dust $\beta$ from the {\it Planck} all-sky dust model. As a corollary, reliable effective temperature maps can be derived which would be otherwise affected by $\beta$ variations. Finally, we note that the angular resolution of extinction maps for the study of nearby starless cores remains unsurpassed.}

\keywords{ISM: dust, extinction, Stars: formation, Submillimeter: ISM, Infrared: ISM, Radio lines: ISM}

   \maketitle
%

\section{Introduction}

The unprecedented wavelength coverage in the far-infrared and submillimeter range provided by {\it Herschel} and {\it Planck} has renewed interest in measuring and constraining the dust properties of the interstellar medium. However, while the data availability has improved tremendously, leading to much better constrained spectral energy distributions, such endeavours are still fundamentally limited by the degeneracy of the dust opacity and temperature in governing the shape of the SED and by the effect of line-of-sight averaging (e.g., \citealp{sch07};\citealp{she09a,she09b};\citealp{kel12}).

Briefly, the specific intensity of optically thin dust emission in the far-infrared and submillimeter wavelength range can be described as $I_\nu = B_\nu(T)\left[1-{\rm e}^{-\tau_\nu}\right] \approx B_\nu(T)\tau_\nu$ where $B_\nu(T)$ is the Planck function that describes blackbody radiation. The frequency dependence of the optical depth $\tau_\nu$ is typically described as $\tau_\nu\propto(\nu/\nu_0)^\beta$. Since $\tau_\nu\propto\kappa_\nu\Sigma_{\rm dust}$, where $\kappa_\nu$ is the dust opacity and $\Sigma_{\rm dust}$ is the dust column density, the spectral index $\beta$ also describes the frequency dependence of the dust opacity $\kappa_\nu\propto(\nu/\nu_0)^\beta$.

In dense cores on scales of $\sim$0.1~pc, the dust opacity spectral index $\beta$ is expected to lie between $\beta=2$, as determined for the diffuse interstellar medium (e.g., \citealp{dra84}), and $\beta\sim1$, as derived for circumstellar disks where grain growth has occurred (e.g., \citealp{bec91}). In dense cores, dust coagulation and grain growth are two out of several processes that are thought to lower $\beta$ from the value that holds for the diffuse ISM. \citet{osh94} modeled the dust emission of protostellar cores and found $\beta\sim1.8$.

Observational evidence for $\beta$ changes in low-mass starless cores has so far remained ambiguous -- largely due to parameter degeneracies. Since these degeneracies are already difficult to disentangle for cores without an internal heating source, we restrict our discussion to starless cores. \citet{kra03} and \citet{bia03}, targeting IC~5146 and Barnard~68, respectively, compared ground-based bolometer data and infrared extinction data to constrain dust opacity ratios. They concluded that a dust opacity spectral index of $\beta\sim2$ is compatible with their data. \citet{shi05} used SCUBA data at 450~$\mu$m and 850~$\mu$m to derive $\beta=2.44\pm0.62$ for the low-mass starless core L1448, assuming a dust temperature of 10.5~K based on radiative transfer calculations. \citet{sch10} observed the starless core TMC-1C with ground-based bolometers and {\it Spitzer} in five different bands; they derive $\beta=2.2\pm0.6$. While these measurements of $\beta$ indicate rather high values, the predictions of \citet{dra84} and \citet{osh94} cannot be ruled out within the uncertainties. Most recently, \citet{sch14} used ground-based millimeter maps to constrain the dust properties of the entire OMC-2/3 region with both starless and protostellar cores. Assuming the dust temperature as equal to the measured kinetic temperature of ammonia, as derived from the NH$_3$(1,1) and (2,2) transitions, they find a very low $\beta=0.9\pm0.3$ across the filament as well as the starless and protostellar cores. 

As part of the {\it Planck} mission, an all-sky model of the Galactic foreground thermal dust emission has been developed; it has been described in detail in  \citet{plaXI}. The model is based on submillimeter data collected at frequencies of 857, 545, and 353~GHz in combination with IRAS 100~$\mu$m data, and it has an angular resolution of 5$'$. The model takes into account three main parameters, i.e., the dust optical depth, the effective dust temperature, and the dust opacity spectral index $\beta$. The effective dust temperature is derived from a modified-blackbody fit toward each line of sight and thus does not directly correspond to a unique physical dust temperature. As explained by \citet{plaXI}, these parameters offer an empirical description of the data, even if their physical interpretation is complicated due to degeneracies and uncertainties. 

After a careful study of parameter interdependencies and the impact of noise and cosmic infrared background anisotropies on the recovered parameters, \citet{plaXI} decided to fit the data in a two-step process. In a first step, to minimize the impact of noise and the cosmic infrared background on the derivation of the dust opacity spectral index $\beta$ and the temperature, $\beta$ is derived on data that has been smoothed to a resolution of 30$'$. Then, in a second step, the temperature and opacity are fitted with a fixed $\beta$ at the nominal resolution of 5$'$. The underlying assumption here is that $\beta$ does not vary strongly on small scales. For the final data product, the typical uncertainty for the fit parameters is estimated at a few percent.

In order to study the relevance of this dust emission model on scales of individual cores, we here present a detailed observational study of the starless core FeSt 1-457 in the Pipe Nebula. Given its nearby location at a distance of 130~pc \citep{lom06}, the Pipe Nebula is an ideal laboratory to study the physics of low-mass starless cores and the initial conditions of star formation (\citealp{lad08,rat08}, \citealp{for09,for10}). The FeSt 1-457 core has been previously targeted with millimeter line observations \citep{agu07,fra10,for14} and deep mid-infrared observations \citep{asc13}. 

In the present study, we use submillimeter imaging and near-infrared extinction mapping to construct maps of the dust column density and the effective dust temperature toward FeSt 1-457. Additionally, we study the correlation of N$_2$H$^+$ emission, as a molecular tracer of cold and dense gas, with the lowest measured effective dust temperatures. In Section~\ref{sec_obs}, we describe the observations before we present the results in Section~\ref{sec_res}. The conclusions are presented in Section~\ref{sec_sum}.

\begin{figure*}
\includegraphics[width=0.5\linewidth]{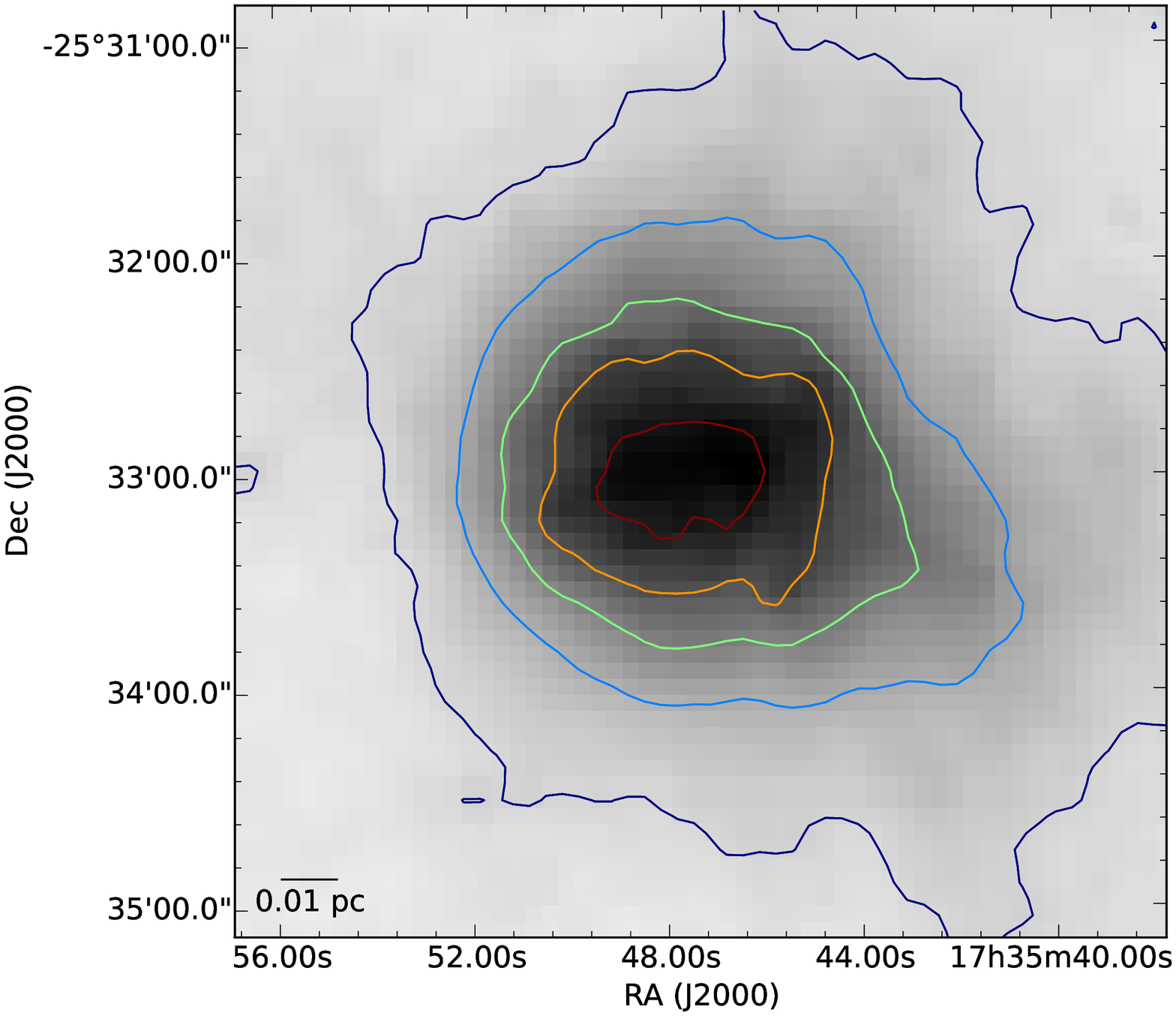}
\includegraphics[width=0.5\linewidth]{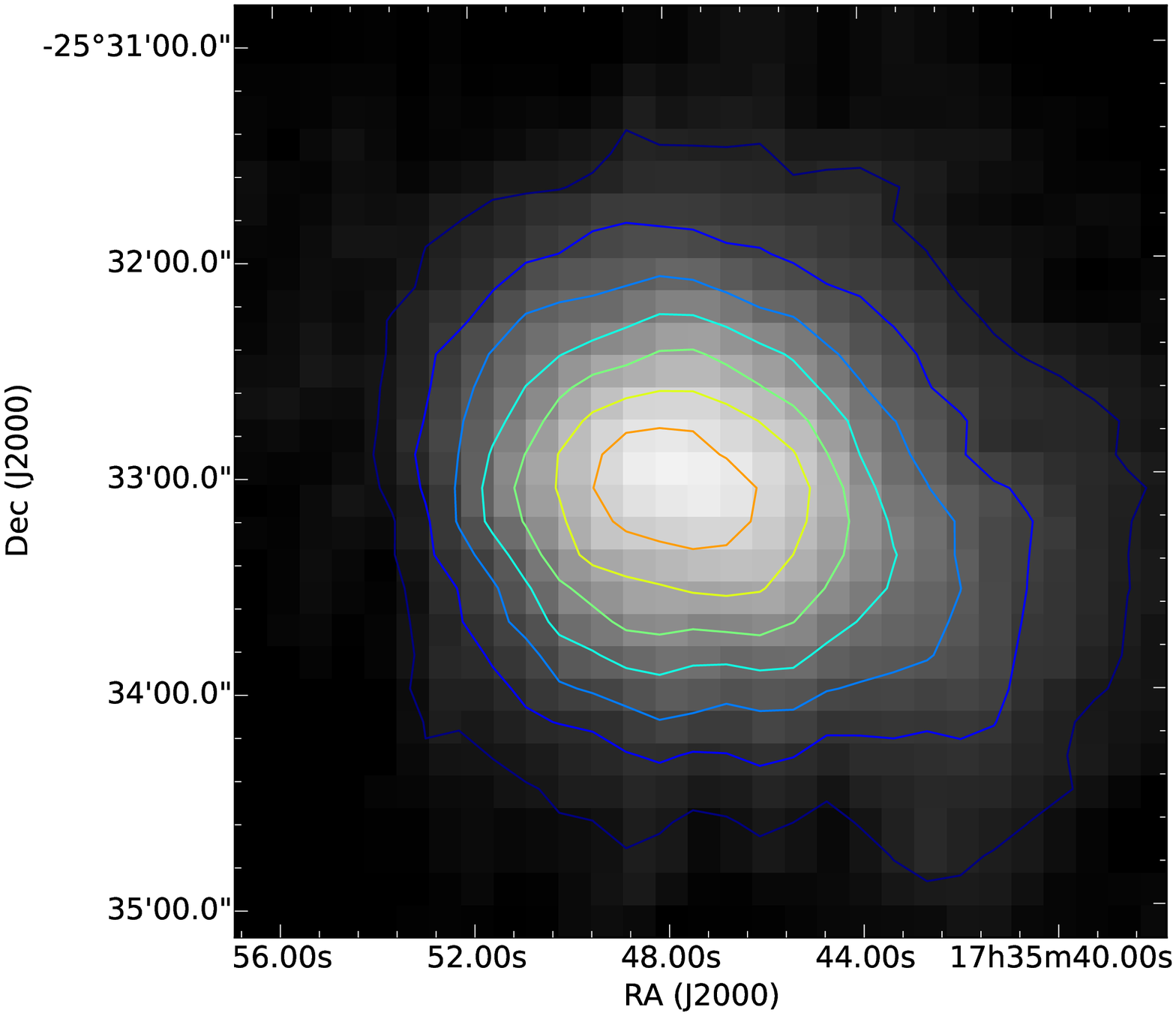}
\caption{Left: Near-infrared extinction map, resolution: 9$''$ FWHM. Contours indicate levels of $A_K$=1 to 5~mag, in steps of 1~mag. Right: {\it Laboca} 870~$\mu$m map, 18$''$ FWHM. Contours indicate levels of 0.033~Jy/beam, corresponding to the 3$\sigma$ limit, to 0.231~Jy/beam, in steps of 0.033~Jy/beam.\label{fig_fest_tauextlaboca}}
\end{figure*}

\section{Observations and data reduction\label{sec_obs}}

\subsection{Herschel submillimeter continuum data} Observations with the {\it Herschel} Spectral and Photometric Imaging Receiver (SPIRE) for this region were obtained on March 14, 2011, as part of a program described by \citet{per12}. The SPIRE detector has three bands with center wavelengths 250, 350, and 500~$\mu$m, or center frequencies 1.2~THz, 857~GHz, and 600~GHz, respectively. These data were processed in combination with {\it Planck} data and 2MASS near-infrared extinction mapping as described by \citet{lom14} to obtain maps of dust optical depth and effective dust temperature at a better angular resolution than from {\it Planck} data alone. This methodology involves the calibration of {\it Herschel} data against the absolutely calibrated {\it Planck} data, and the use of 2MASS extinction maps to convert dust optical depth to column densities. Our analysis generally uses the largest {\it Herschel} beam size of 36$''$, which is an improvement in angular resolution by a factor of $\sim$8 when compared to the corresponding {\it Planck} data. 

Following the procedure developed by \citet{lom14}, we created maps of effective dust temperature from the calibrated {\it Herschel} data. To describe the dust emission along each line of sight and thus derive these effective dust temperatures, we have fitted a modified blackbody to the long-wavelength {\it Herschel} spectral energy distribution (SED), using a dust opacity spectral index $\beta$ fixed to the value determined in the {\it Planck} all-sky model. A single effective dust temperature, as a weighted average of temperatures along each line of sight, is sufficient to fit the data, but since the temperature can vary along the line of sight and still produce a well-behaved effective dust temperature, it does not mean that isothermal conditions are assumed. Due to these complications, the effective dust temperature does not directly correspond to a physical dust temperature of strictly isothermal dust even though it indirectly reflects a weighted average of such dust temperatures along the line of sight. 

\subsection{APEX-Laboca submillimeter continuum data\label{sec_teclaboca}}
We additionally obtained ground-based submillimeter data of the FeSt 1-457 core and its surroundings with the Large APEX Bolometer Camera ({\it Laboca}). {\it Laboca} is a submillimeter bolometer array \citep{sir09} and a facility instrument of the Atacama Pathfinder Experiment (APEX) telescope \citep{gus06}. The telescope is located on Llano de Chajnantor in the Chilean Andes, at an altitude of 5107\,m. The central frequency of the {\it Laboca} bolometers is 345~GHz (870~$\mu$m), with a spectral passband that is about 60~GHz, or 150~$\mu$m, wide (FWHM).  Observations were carried out as part of our project 84.C-0927 on 23 and 24 September 2009. The FeSt 1-457 core was mapped for about 35 minutes with fluxes calibrated against the primary calibrator Neptune (and G5.89). A combination of spiral maps and on-the-fly maps was used to cover the area.  At the {\it Laboca} central wavelength, the beam size of the APEX telescope is 19$''$ FWHM. The data have been processed according to the iterative procedure described by \citet{bel11} to maximize S/N. The final noise level from combining the individual maps is 11~mJy/beam. In our analysis, we take into account the theoretical instrument spectral response as published in \citet{sir09} and on the APEX website\footnote{\url{http://www.apex-telescope.org/bolometer/laboca/technical/}}.

\subsection{Near-infrared data\label{sec_nir}}
In addition to the 2MASS data that are used as part of the {\it Herschel} calilbration procedure described above, we use deeper near-infrared data to obtain an even better extinction map of FeSt 1-457. Since its angular resolution exceeds that of {\it Herschel} and {\it Laboca}, this map serves as a reference comparison for both.
The near-infrared ground-based observations used in this context are part of a
high-resolution near-infrared survey of regions across the whole Pipe Nebula
\citep{rom09,rom10}. The data corresponding to the Fest 1-457
core, come from two data sets. The first set is of $H$ and $K_s$ images
obtained with the Son of ISAAC (SofI) camera at the ESO New Technology
Telescope (NTT) atop Cerro La Silla in Chile. The resolution of the 
SofI-NTT observations is $0\farcs288$. The second set of ground-based 
data are $H$ and $K_s$ images obtained with the Infrared Spectrometer 
And Array Camera (ISAAC) at the ESO Very Large Telescope (VLT) atop Cerro Paranal in Chile. These ISAAC-VLT observations were carried out during 2002 July. The VLT observations targeted the areas of peak extinction of the
core and its nucleus. The high angular resolution ($0\farcs148$) and
sensitivity of these observations allowed us to detect several
highly reddened background sources that could not be detected with SofI,
allowing us to construct a complete NICEST extinction map \citep{lom09} of the core at a resolution of 9$''$ (FWHM). At this resolution and a pixel size of $4\farcs5$, it is still possible to ensure the presence of a minimum of about ten (and often many more) background stars per resolution element (beam) for this method to work.

\subsection{Millimeter line data} Observations in N$_2$H$^+$(1-0) and other transitions were obtained with the IRAM 30m telescope in July 2003. These observations have been reported and described in detail by \citet{agu07}, and we here use their N$_2$H$^+$(1-0) map to compare this tracer of cold and dense gas with our effective dust temperature map.

\begin{figure*}
\includegraphics[width=0.5\linewidth]{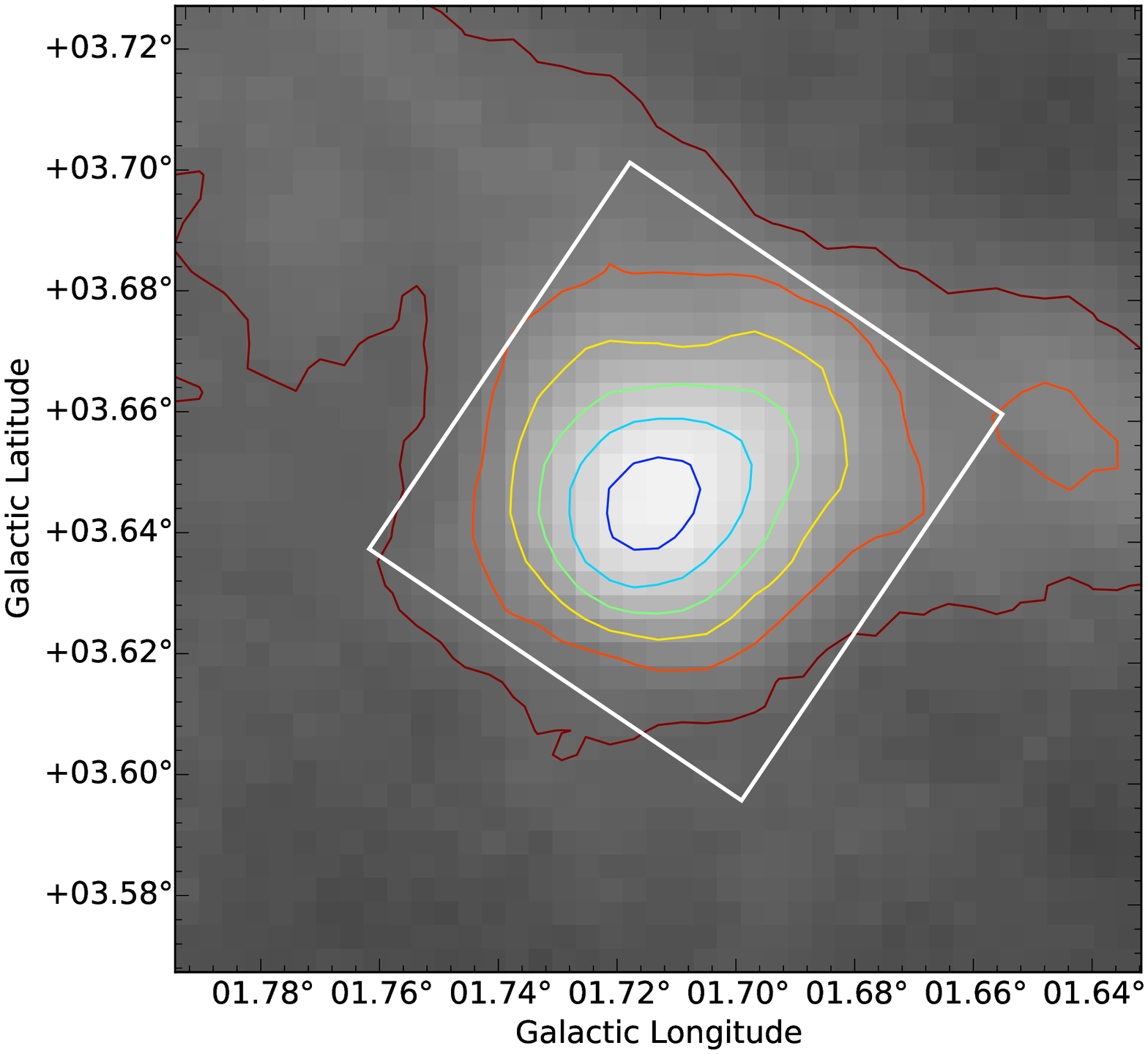}
\includegraphics[width=0.5\linewidth]{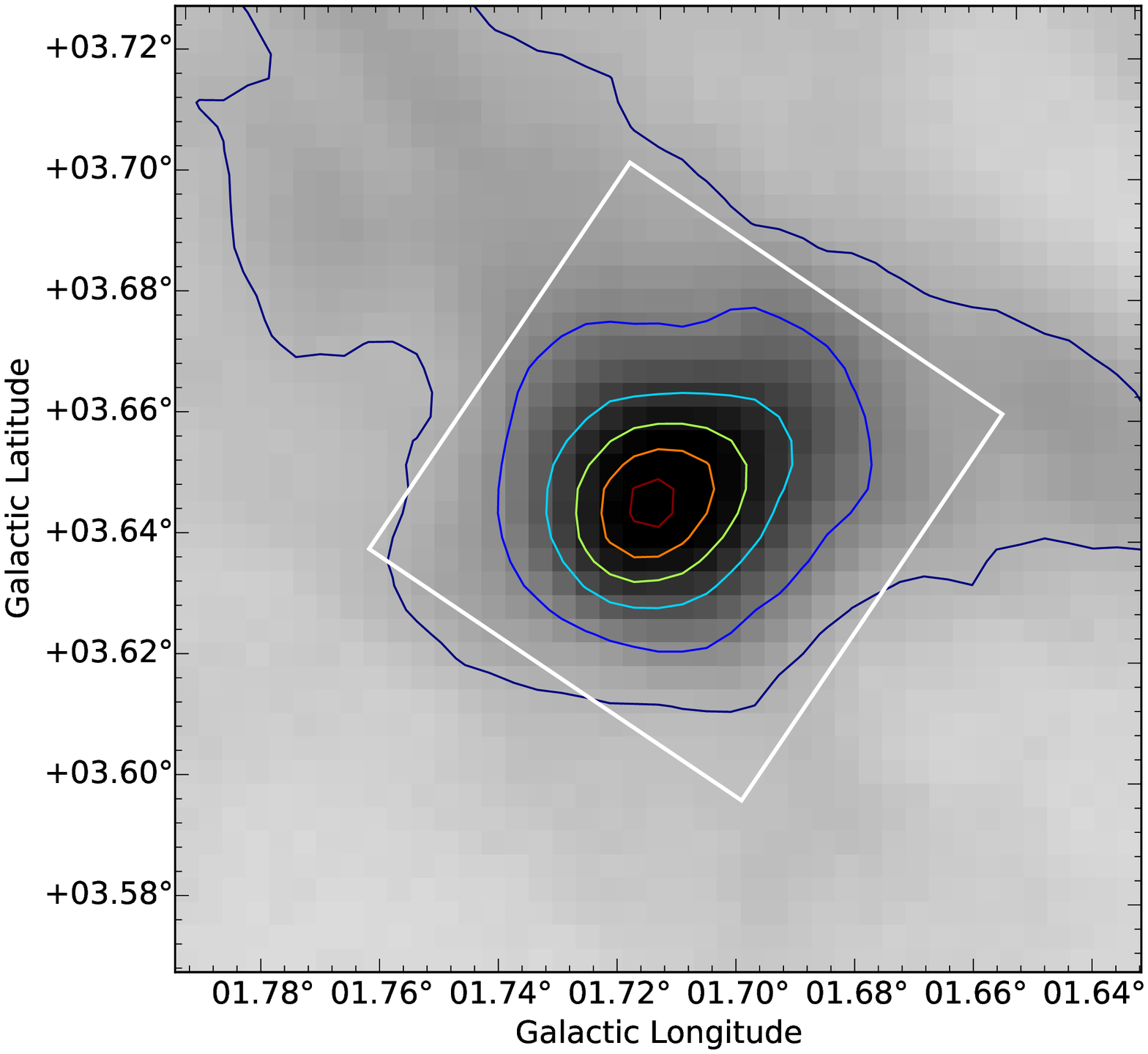}
\caption{Left: Effective dust temperature map of FeSt~1-457, 36$''$ FWHM, with contours indicating temperatures of 20~K down to 14~K in steps of 1~K. Right: Corresponding optical depth map, with contours indicating levels of 2, 3, 5, 7, 9, and 11$\times 10^{-4}$. In both panels, the white box indicates the field of view of Figure~\ref{fig_fest_tauextlaboca}.\label{fig_fest_tautemp}}
\end{figure*}

\section{Results and Analysis\label{sec_res}}

For a general overview of the FeSt 1-457 core, we show the high-resolution extinction map (9$''$ FWHM) and the {\it Laboca} dust emission map (18$''$ FWHM) in Figure~\ref{fig_fest_tauextlaboca}. The {\it Herschel}-based effective dust temperature and dust optical depth maps are shown in Figure~\ref{fig_fest_tautemp}. We use the best available angular resolution wherever possible, but to compare all three datasets, we use a common resolution of 36$''$, reached by convolving the higher-resolution data to larger beam sizes. At a distance of 130~pc, an angular size of 36$''$ corresponds to a size scale of 0.02~pc. For intercomparisons, the extinction and {\it Laboca} data were thus convolved to this lower resolution.

In the following, we first use the {\it Herschel} SED fit and resulting dust model to demonstrate that the FeSt~1-457 core is not isothermal. Then, we compare the {\it Herschel}-derived dust optical depth with the independently derived near-infrared extinction map to constrain the dust opacity $\kappa_\nu$. Finally, we use the {\it Herschel} dust model to predict the submillimeter flux density observed by {\it Laboca}. Confronting this prediction with the observations allows us to constrain the dust opacity spectral index $\beta$.

\subsection{Temperature profile}

\begin{figure}
\centering
\includegraphics[angle=0, width=1.00\linewidth]{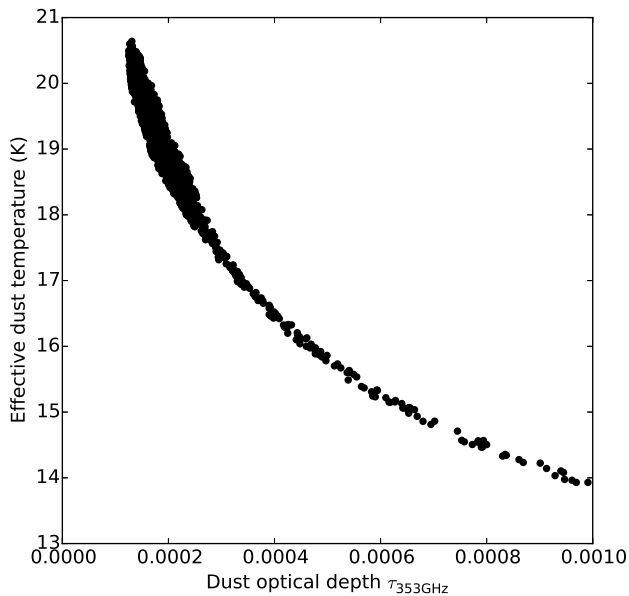}
\caption{Correlation of the submillimeter dust optical depth (at 353~GHz) and the effective dust temperature, as derived using the method described by \citet{lom14}, at a resolution of 36$''$ FWHM.\label{fig_fest_tau_T}}
\end{figure}

\begin{figure}
\centering
\includegraphics*[angle=0, width=1.00\linewidth]{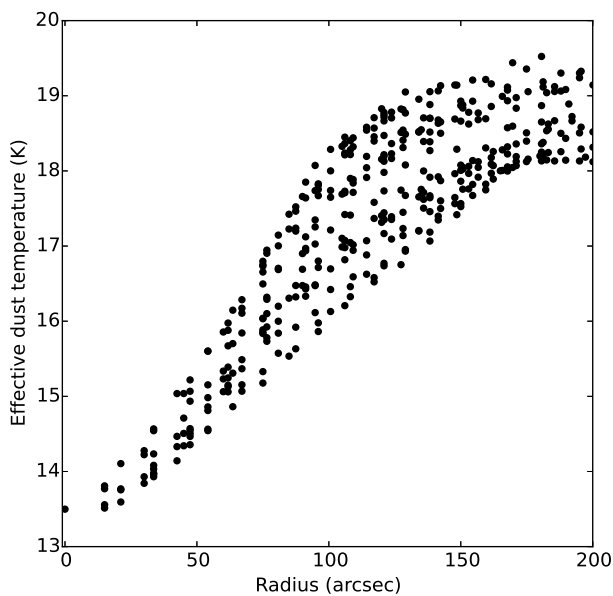}
\caption{Effective dust temperature toward the FeSt 1-457 core, as a function of core radius.\label{fig_temp_profiles}}
\end{figure}

At a resolution of 36$''$ FWHM, the analysis of the {\it Herschel} effective dust temperature and 353~GHz dust optical depth, as derived according to \citet{lom14}, reveals significant gradients in both quantities and a tight anticorrelation, shown in Figure~\ref{fig_fest_tau_T}. The FeSt~1-457 core shows a clear decrease in effective dust temperature toward its center. A plot of the same temperature data versus radius of the core is shown in Figure~\ref{fig_temp_profiles}. The lowest effective dust temperature at the center of the core is 13.50$\pm$0.05~K. This is 4~K higher than the kinetic temperature of $T_K=9.5\pm0.1$~K derived from dense-gas--tracing NH$_3$ observations obtained with the Green Bank Telescope and a corresponding beam size of 30$''$ \citep{rat08}. 

However, in the {\it Herschel} data, we do not only see the gas and dust at the center of the core. Instead, we always observe the entire line of sight through the core with its entire temperature structure. The effective dust temperature, as a weighted average along the line of sight derived from an SED fit, is our main single observable of the temperature conditions in the core. Similarly, when we use the same SED model to predict the flux density as observed with {\it Laboca}, it is the same effective dust temperature that enters the Planck function consistently. Deconvolving the actual absolute temperature profile from the observed effective dust temperature profile would require detailed modeling of the core which is beyond the scope of this paper.

If the density is sufficiently high, we would expect the gas and dust to be thermodynamically coupled such that the dust temperature and the gas kinetic temperature would be the same, but it is not obvious whether and where that would be the case in the FeSt 1-457 core. Assuming that the NH$_3$ kinetic temperature is a good approximation of the dust temperature at the center of the core would indicate that the effective dust temperature toward the center overestimates the actual dust temperature at the center by about 4~K which is similar to findings reported for Barnard~68 (see Section~\ref{sec_b68}).

We can shed additional light on the gas-to-dust coupling by comparing the effective dust temperatures to molecular emission of N$_2$H$^+$ as a tracer of high volume densities. When studying a large sample of Pipe cores, including the FeSt 1-457 core, \citet{for14} noted a correlation of N$_2$H$^+$(1-0) detections with low effective dust temperatures. Out of 52 cores, only the coldest six, with effective dust temperatures between 13.2 and 14.6~K were detected in N$_2$H$^+$ emission, a tracer of high volume densities. While that study only contained a single Mopra pointing at the FeSt 1-457 core (source 96), it was among the N$_2$H$^+$ detections. 

\begin{figure}
\centering
\includegraphics*[angle=0, width=1.00\linewidth]{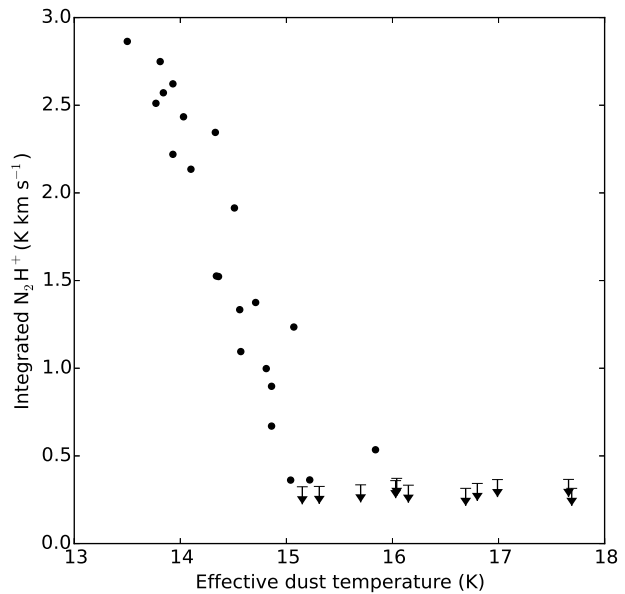}
\caption{Integrated N$_2$H$^+$ emission (including all HFS components) vs effective dust temperature in the FeSt 1-457 core. Arrows denote 3$\sigma$ upper limits. The emission is clearly correlated with the coldest temperatures.\label{fig_T_n2hp_int}}
\end{figure}

We can now look at the correlation of effective dust temperatures and N$_2$H$^+$ emission in more detail by making use of the N$_2$H$^+$(1-0) map of FeSt 1-457 that was obtained by \citet{agu07}. Using the IRAM 30m telescope, they reach an angular resolution of 26$''$ FWHM. We use the same data that have been described in \citet{agu07}, and we simply integrate the spectra over velocity intervals that cover the N$_2$H$^+$(1-0) hyperfine structure components. To approximately compare with the {\it Herschel} effective dust temperatures, we have read out the closest pixel from the column density map with 36$''$ FWHM; the pixel size is 15$''$. The result is shown in Figure~\ref{fig_T_n2hp_int}. The strongest emission correlates with the lowest effective dust temperature, and there is a steep fall-off with rising temperatures, similar to the temperature range with N$_2$H$^+$ detections that was reported by \citet{for14}. Interestingly, the correlation of N$_2$H$^+$ emission and low effective dust temperatures thus is not only seen in a sample of single observations of many different cores, but also in a map of a single core. Even as a weighted average along the line of sight, the effective dust temperatures can serve as a means of identification of the cold and dense conditions inside molecular cores.

\subsection{The dust opacity $\kappa_\nu$}

A comparison of the dust optical depth $\tau$ with the independently derived near-infrared extinction yields a contraint on changes of the dust opacity $\kappa_\nu$. When studying regions of comparably low extinction ($A_K<2$~mag), \citet{lom14} find a linear relation $A_K=\gamma\tau_{353{\rm GHz}}+\delta$ where $\gamma$ is proportional to the ratio of the near-infrared extinction coefficient at 2.2~$\mu$m and the submillimeter dust opacity $\kappa_{353{\rm GHz}}$, $\gamma\propto C_{2.2}/\kappa_{353{\rm GHz}}$. In our case, the slope of $\tau_{353{\rm GHz}}(A_K)$ is proportional to $1/\gamma$.

\begin{figure}
\centering
\includegraphics[angle=0, width=1.00\linewidth]{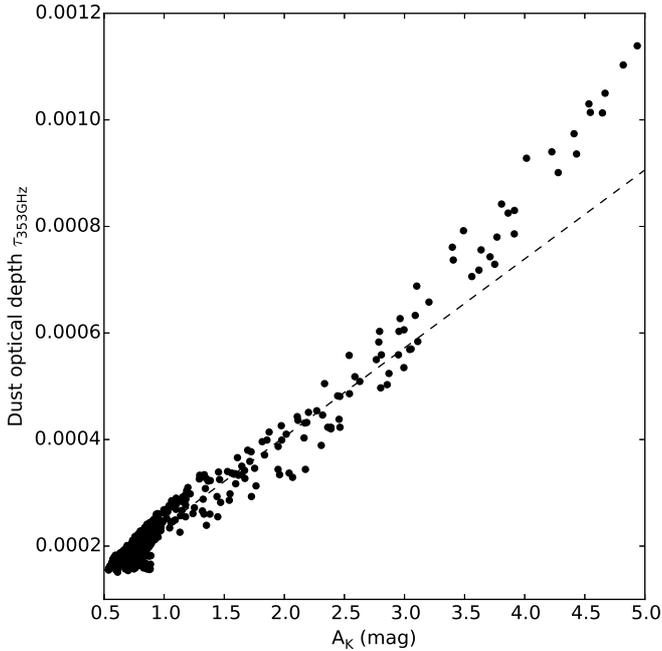}
\caption{Comparison of dust optical depth $\tau_{\rm 353~GHz}$ and near-infrared extinction $A_K$. The dashed line shows a linear fit of the points with $A_K<2$~mag. \label{fig_tauak36}}
\end{figure}

In Figure~\ref{fig_tauak36}, we show a comparison of the dust optical depth and an extinction map, produced with matching 36$''$ FWHM resolution. The correlation between the two quantities is clearly nonlinear. For reference, we show a linear fit to the low-extinction part of the plot, only taking into account points with $A_K<2$~mag. Toward the highest near-infrared extinctions, above approximately $A_K\sim2$~mag, the submillimeter dust optical depth $\tau_{\rm 353 \rm{GHz}}$ rises faster than predicted from this low-extinction linear fit. To stay in the picture introduced by \citet{lom14}, this means that $1/\gamma$ increases with extinction.

An increase of $1/\gamma$ toward the highest levels of extinction, i.e., an increase toward the center of the core, means that either the submillimeter dust opacity $\tau_{\rm 353 \rm{GHz}}$ is increasing or the near-infrared extinction coefficient is decreasing, or both. If grain growth occurs in the central regions of this core, this would exactly be the expected observational signature (e.g., \citealp{orm11}). In a study of the mid-infrared extinction law of the same core, \citet{asc13} had found evidence for large grains in the very same core (FeSt 1-457), but throughout the entire core and not just its innermost regions. Since the sensitivity to given grain sizes is different between these various methods, the finding of \citet{asc13} may well be compatible with our finding. In combination, this may mean that in the inner core, it is mostly the submillimeter dust opacity that is changing and not the infrared extinction coefficient.

We can make a simple estimate of the volume densities occurring in the central regions of the FeSt~1-457 core by determining the average volume density along the line of sight under the assumption of spherical symmetry. The peak extinction is $A_K=4.88$~mag on a size scale of 0.02~pc. Assuming an approximate conversion of $N_H\sim2\times10^{22}A_K$~cm$^{-2}$ \citep{ryt96,vuo03}, this extinction corresponds to a column density of $9.76\times10^{22}$~cm$^{-2}$. Given a projected extent in the plane of the sky of 240$''$ or 0.15~pc, this column density corresponds to an average volume density of $2.1\times10^{5}$~cm$^{-3}$. This is a lower limit for the volume density at the center of the core but it is nevertheless an estimate for the volume densities where we find evidence for grain growth. 

\subsection{The dust opacity spectral index $\beta$}

In addition to the {\it Herschel} dust model and the near-infrared extinction map, we can now use the {\it Laboca} submillimeter map to derive further constraints on the dust emission. We argue that in particular it is possible to constrain the dust opacity spectral index $\beta$. 

\begin{figure}
\centering
\includegraphics[angle=0, width=1.00\linewidth]{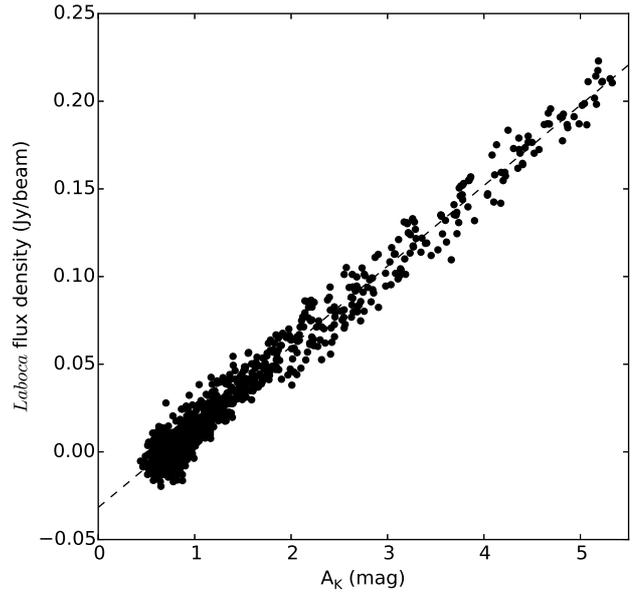}
\caption{Comparison of the {\it Laboca} map and a NICEST near-infrared extinction map with matching resolution (18$''$ FWHM). The dashed line shows a linear fit above $A_K>1$~mag to avoid noisy points.\label{fig_laboca_vlt_comp}}
\end{figure}

First of all, in a measurement that is entirely independent of the {\it Herschel} dust model, we find a {\it linear} trend of the {\it Laboca} submillimeter flux density as a function of near-infrared extinction $A_K$. In Figure~\ref{fig_laboca_vlt_comp} we show this trend at the native {\it Laboca} resolution of 18$''$ FWHM, using an extinction map with the same resolution. The dashed line shows a simple linear fit ($f(x)=ax+b$) to the data where $a=0.0459\pm0.0003$ and $b=-0.0317\pm0.0008$~Jy/beam (18$''$ FWHM). A zeropoint offset becomes apparent that we will quantify and discuss below for data convolved to 36$''$ FWHM. To within the errors, the comparison of {\it Laboca} submillimeter emission and near-infrared extinction reveals a linear correlation that holds up to the highest extinction levels. This observational result is surprising since, as we will discuss now, the flux density depends on both the dust optical depth and the temperature, and both are nonlinear functions of $A_K$, as we have seen.

It turns out that {\it Laboca} allows us to separately probe the region for evidence of a change in the dust opacity spectral index $\beta$. For this purpose, we use the {\it Herschel} dust model to predict the flux density observed by {\it Laboca} to then confront this prediction with the observations. To match the resolution of the dust model, the {\it Laboca} data have been convolved to match the resolution of the {\it Herschel} data at 36$''$ FWHM. 

Assuming optically thin submillimeter dust emission, which is a safe assumption judging by the {\it Herschel} data, the predicted {\it Laboca} flux density for every line of sight depends on the effective dust temperature, the dust optical depth $\tau_\nu$, and the dust optical depth (and opacity) spectral index $\beta$: $S_\nu = B_\nu(T)\cdot(1-e^{-\tau_\nu})\cdot\Omega_b\cdot10^{26}$ Jy/beam, where $B_\nu(T)$ is the Planck function and $\Omega_b$ is the beam solid angle. To take into account the {\it Laboca} spectral response, described in Section~\ref{sec_teclaboca}, we determine $S_\nu$ for a list of frequencies across the {\it Laboca} spectral response and weigh these individual flux densities into a total flux density according to their relative weights. Note that the {\it Laboca} fluxes are implicitly calibrated against the spectrum of Neptune with a spectral slope of about $\beta_{\rm planetary}\sim 2$, which is close to the spectral slopes discussed here, so that no color correction is applied (see also \citealp{plaIX}). Such a correction would be by less than one percent in flux density which is an accuracy anyway not reached by ground-based submillimeter measurements.

The only two variables in this prediction are the temperature $T$ and the dust opacity spectral index $\beta$, used to calculate the dust optical depth for every frequency in the {\it Laboca} pass band, starting from $\tau_{\rm 353 GHz}$, where the reference frequency already lies inside the {\it Laboca} pass band. For these two variables, $T$ and $\beta$, we use the effective dust temperature and the dust opacity spectral index from the Planck dust model at this position, $\beta_{\rm Planck}=1.68\pm0.06$. While the effective dust temperature has already been derived from a Planck function, the choice of the Planck-$\beta$ is less obvious even if there is no alternative due to the lack of other constraints. In the {\it Herschel} dust model, the Planck-$\beta$ has already been used to determine $\tau_{\rm 353 GHz}$, extrapolating from a frequency near the peak of the SED (here assumed to be 857~GHz or 350~$\mu$m). Since the Planck-$\beta$ has been derived on scales much larger than the core (30$'$, or 1.1~pc at a distance of 130~pc), its relevance for the description of the dust emission on (sub-)core scales is a priori unclear.

\begin{figure}
\includegraphics*[width=\linewidth]{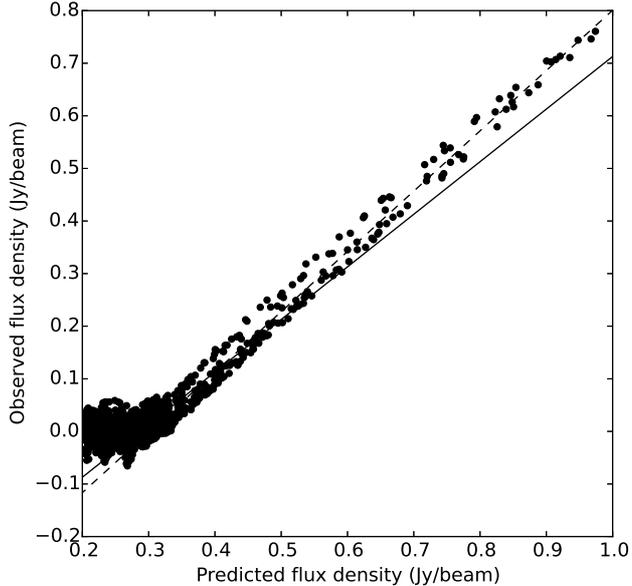}
\caption{Comparison of the observed submillimeter flux density from {\it Laboca} and the predicted flux density based on the {\it Herschel}/{\it Planck} model. The dashed line shows a linear fit of all points with a predicted flux density of $>$0.4 Jy/beam. For comparison, the continuous line indicates the reference slope of 1 (arbitrarily set to intersect with the fit at its starting point of $x=$0.4 Jy/beam) which would correspond to $\beta=\beta_{\rm Planck}$. The difference in slope corresponds to a difference of $\Delta\beta=0.15$ (see text).\label{fig_sobs_spred}}
\end{figure}

In Figure~\ref{fig_sobs_spred}, we directly compare the submillimeter flux density, as observed with {\it Laboca}, with the flux density that is predicted based on the {\it Herschel} dust model. The direct comparison of observed vs. predicted flux density, taking into account the temperature profile of the core in particular, results in a linear relation. There is an obvious zeropoint offset between the observed and predicted flux densities, which we will discuss below. Also, the systematic pattern of points in the plot, which look like two separate branches of the correlation, can be explained by a subpixel relative shift in astrometry between the {\it Herschel}/{\it Planck} and {\it Laboca} maps which cannot be verified independently. The slope of the relation is close to one which is remarkable, given that the local dust $\beta$ may have nothing to do with the one derived from {\it Planck} data on larger scales of 30$'$. 

According to a linear fit, the {\it Laboca} flux density at the {\it Herschel}/{\it Planck} zeropoint is $-0.346\pm0.002$ Jy/beam. The fit result for the slope is $a=1.147\pm0.006$ up to the highest flux densities, corresponding to the highest column densities. This means that a single dust $\beta$ is sufficient to describe these data. However, the fact that the slope is different from one indicates that the true value of $\beta$ is slightly different from $\beta_{\rm Planck}$. Since the dust model is able to predict the observed submillimeter flux densities, apart from zeropoint offsets, this result also means that changes in the dust optical depth and the effective dust temperature partly cancel out in the observed flux density to produce a linear relation of flux density vs. extinction.

In order to test whether the {\it Laboca} measurements can be explained by a single but potentially different $\beta_{\rm N}\neq\beta_{\rm Planck}$ we first use $\beta_{\rm Planck}$ to obtain the optical depth at a {\it Herschel} frequency near the peak of the SED (here assumed to be 857~GHz or 350~$\mu$m) by extrapolation from $\tau_{\rm 353GHz}$. Then we introduce a different $\beta_{\rm N}$ as a variable to extrapolate back to the {\it Laboca} observing frequency, such that the combined application yields the measured {\it Laboca} flux density. The ratio of measured to predicted flux density then is:

\begin{equation}
\frac{S_{\rm Laboca}}{S_{\rm predicted}} = \left(\frac{857 {\rm GHz}}{{353 {\rm GHz}}}\right)^{\beta_{\rm Planck}-\beta_{\rm N}}
\end{equation}

The slope of the linear function in Figure~\ref{fig_sobs_spred} thus is a direct measure of the $\Delta\beta = \beta_{\rm Planck}-\beta_{\rm N}$, where $\beta_{\rm N}$ provides a better description of the data: a ratio of the observed vs predicted flux densities yields $(857/353)^{\Delta\beta} = a$. We then find $\Delta\beta = {\rm ln}(a)/{\rm ln}(857/353) = 0.15\pm0.01$ where the error estimate merely propagates the error in $a$ from the fit. The result is that a dust $\beta$ that is close to the {\it Planck} value can describe the data: $\beta_{\rm N} = 1.68-0.15 = 1.53\pm0.07$. In fact, $\beta_{\rm N}$ is compatible with the {\it Planck} result to within the nominal respective 2$\sigma$ errors. It may be worth pointing out that even the use of the improved $\beta_{\rm N}$ results in an underprediction of the highest observed flux densities, i.e., those at the center of the core.

The advantage of this estimate is that it does not depend on zeropoint offset that is clearly present in the data shown in Figure~\ref{fig_sobs_spred}. Two effects contribute to this zeropoint offset: any actual calibration offset between the two datasets and spatial filtering of large-scale emission in the {\it Laboca} data. Additionally, the fact that the relation of measured to predicted flux density is entirely linear with a zeropoint offset suggests that spatial filtering in the {\it Laboca} data due to correlated-noise removal does not set in at angular scales that would affect the core. \citet{bel11} show that the ratio of output to input flux density due to spatial filtering in {\it Laboca} data for a Gaussian source of size 200$''$ (FWHM) is still 90\%. Our source FeSt 1-457 is smaller than this size, and the spatial filtering in this case only affects a constant background flux density, corresponding to the largest angular scales. Indeed, the {\it Laboca} map falls to zero flux density outside of the core while the {\it Herschel}/{\it Planck} map does not. We thus have a situation where a zeropoint offset is caused both by a possible calibration offset and spatial filtering on the largest scales. 

Alternatively, to reconcile the {\it Laboca} measurements and the {\it Herschel}/{\it Planck} prediction, we could leave the dust $\beta$ unchanged and instead modify the effective dust temperatures to match the predicted and observed flux densities. Even if in reality both the dust opacity spectral index $\beta$ {\it and} the temperatures differ from our assumptions, changing just one of the two provides us with two instructive thought experiments. 

After concluding that a single dust opacity spectral index $\beta_{\rm N}$ that is only slightly different from the {\it Planck}-derived value can reconcile the {\it Laboca} and {\it Herschel}/{\it Planck} datasets, we now solve $S_\nu = B_\nu(T_d)\tau_d\Omega$ for the dust temperature $T_d$ while requiring the submillimeter flux density to match the observed value. In this experiment, it is mandatory to take into account the zeropoint offset. If both the different slope and the offset are to be explained by temperature effects, the result is an unphysical temperature profile for a starless core where the temperature would rise toward the center. Since we know that the {\it Laboca} measurements may have an effective zeropoint offset due to spatial filtering, we solve for a temperature prediction after adding 0.346~Jy/beam to every pixel, as derived from the linear fit discussed above. 

Without the zeropoint offset, generally {\it higher} effective dust temperatures could account for the difference between the measured and predicted {\it Laboca} fluxes. The temperatures change in nonlinear fashion, and the scatter for a given core radius increases considerably (not shown). While this scenario of individually modified temperatures cannot be ruled out, we prefer the scenario with a single, if slightly modified dust-$\beta$ because of its simplicity. Higher temperatures do not appear to be plausible since the line-of-sight weighted averages of the dust temperature are already an overestimate of the local dust temperature in the inner core.

In the light of the recently reported results of \citet{sch14}, we note that we have also tried to emulate their experiment for the case of FeSt 1-457. However, we do not have a map of the gas kinetic temperature of NH$_3$, but only a single measurement toward the center of the core, as mentioned above. As a zero-order approximation, we have subtracted the difference between the minimum effective dust temperature and the ammonia gas kinetic temperature, a difference of 4~K, from all points of our effective temperature map, thereby forcing the observed dust temperature toward the center of the core to be equal to the ammonia gas kinetic temperature while keeping the temperature structure of the map. We find that repeating the above experiment then requires a very different $\beta=0.64$ to explain the data. While this value would be within the range reported by \citet{sch14}, it would be very different from the {\it Planck} value. However, given that we know that a temperature gradient is present in such cores and given the degeneracy of $\beta$ and $T$ in the observations and its impact on the effective dust temperature as an observable, as well as uncertainties as to when the gas and dust can be assumed to be fully thermodynamically coupled, it is not obvious that any gas kinetic temperature can directly be used as a physical dust temperature in the line of argument described above.

\subsection{Zeropoint offsets}

This leaves the zeropoint offset between {\it Herschel}/{\it Planck} and {\it Laboca} to be explained. As determined from the fit, the {\it Laboca} flux density at the {\it Herschel}/{\it Planck} zeropoint is $-0.346\pm0.002$ Jy/beam. As noted above, the most obvious reason for this offset lies in the spatial filtering of large-scale structure, larger than the FeSt 1-457 core. Indeed, at a distance of about $2\farcm5$ from the core, the {\it Laboca} flux density falls to zero while the {\it Planck} never reach zero in the neighborhood of the FeSt 1-457 core. We directly compare the {\it Laboca} map to the {\it Planck} image at 353~GHz, i.e., at essentially the same observing frequency as {\it Laboca}. With a beam size of about $4\farcm8$ FWHM, the {\it Planck} data have lower angular resolution compared to {\it Laboca}, but they cover the entire area. Selecting an image of 17$'\times$17$'$ (containing 22.1 FWHM beam areas), the contrast of the peak, at the position of the core, to the mean pixel value of the image is only 1.38. If we assume that this contrast represents the large-scale emission, this would mean that at the peak of the {\it Laboca} image, where a flux density of 0.77~Jy/beam (36$''$ FWHM) is measured, extended emission of 0.29~Jy/beam has been filtered out which is very close to the observed zeropoint offset.

This suggests that a large fraction of extended emission, unrelated to the core, is filtered out by the {\it Laboca} imaging, but it is present in the {\it Planck} data. The {\it Planck} point source catalog provides us with an additional means to judge the spatial filtering. The Planck Catalog of Compact Sources Release 1 \citep{plaXXVIII} lists the FeSt 1-457 core as G001.72+03.64. The catalog lists aperture photometry where the aperture has a radius equal to the FWHM beam size and an adjacent annulus for local background subtraction. The annulus has an outer radius of 2$\times$FWHM. At 353~GHz, this corresponds to an aperture radius of about $4\farcm8$ and a radius of the outer annulus of $9\farcm6$. This corresponds well to the total field of view (12$'$) of the entire {\it Laboca} bolometer array and should thus give a first-order approximation of the large-scale spatial filtering. The flux density (APERFLUX) is catalogued as $8.248\pm3.334$~Jy. Corresponding PSF-fitting photometry (which also takes into account a background offset) is listed as $8.570\pm1.214$~Jy. Applying the same aperture on the {\it Laboca} data yields a total flux density of 8.034~Jy which is in good agreement with the {\it Planck} flux. We note that in the {\it Laboca} data, the background annulus contains flux that is indistinguishable from zero flux, and the entire emission of the core can be contained in a significantly smaller aperture than the one used for {\it Planck}.

\subsection{Higher resolution: Extinction mapping}

\begin{figure*}
\centering
\includegraphics*[angle=0, width=0.45\linewidth]{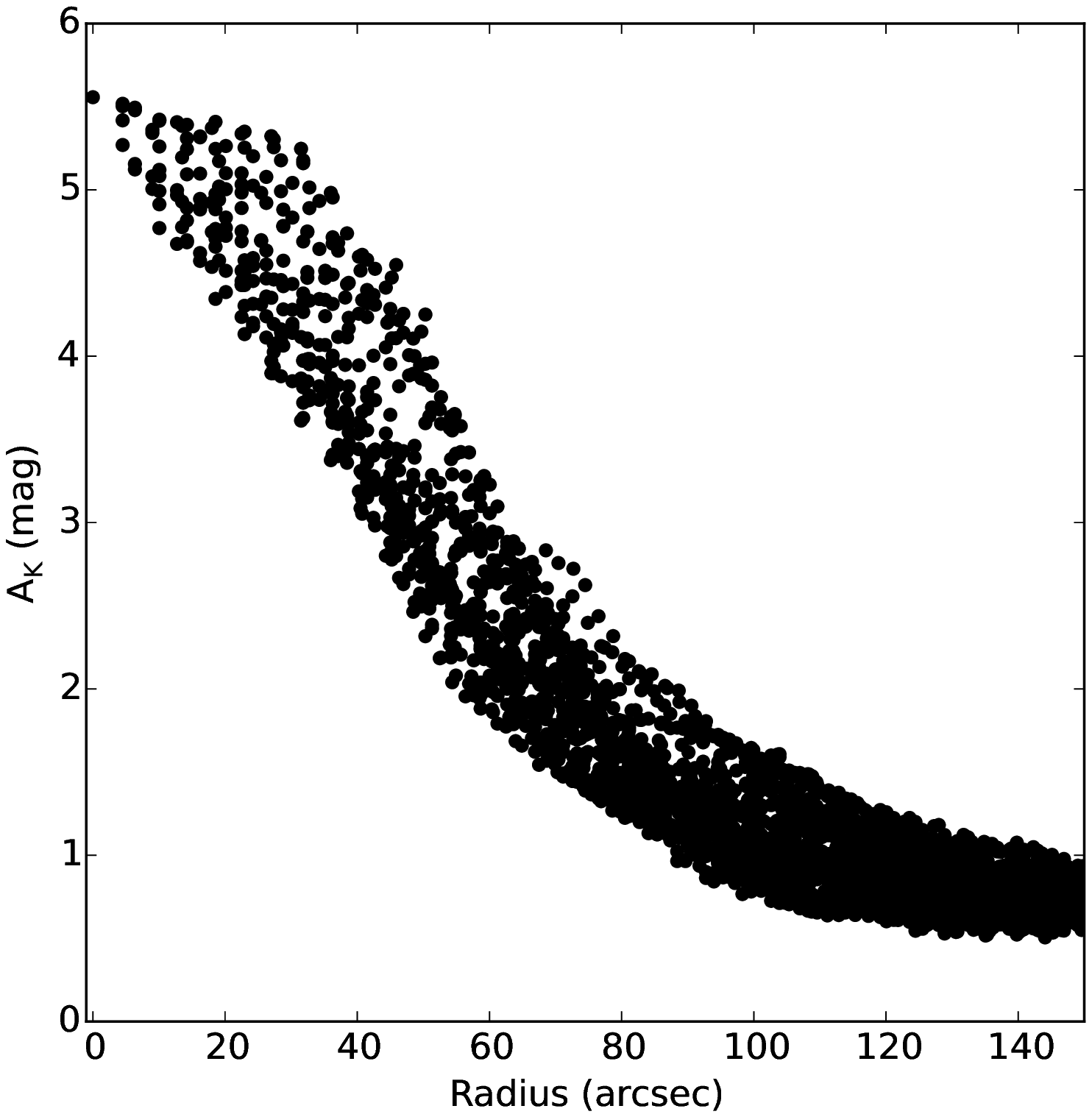}
\includegraphics*[angle=0, width=0.45\linewidth]{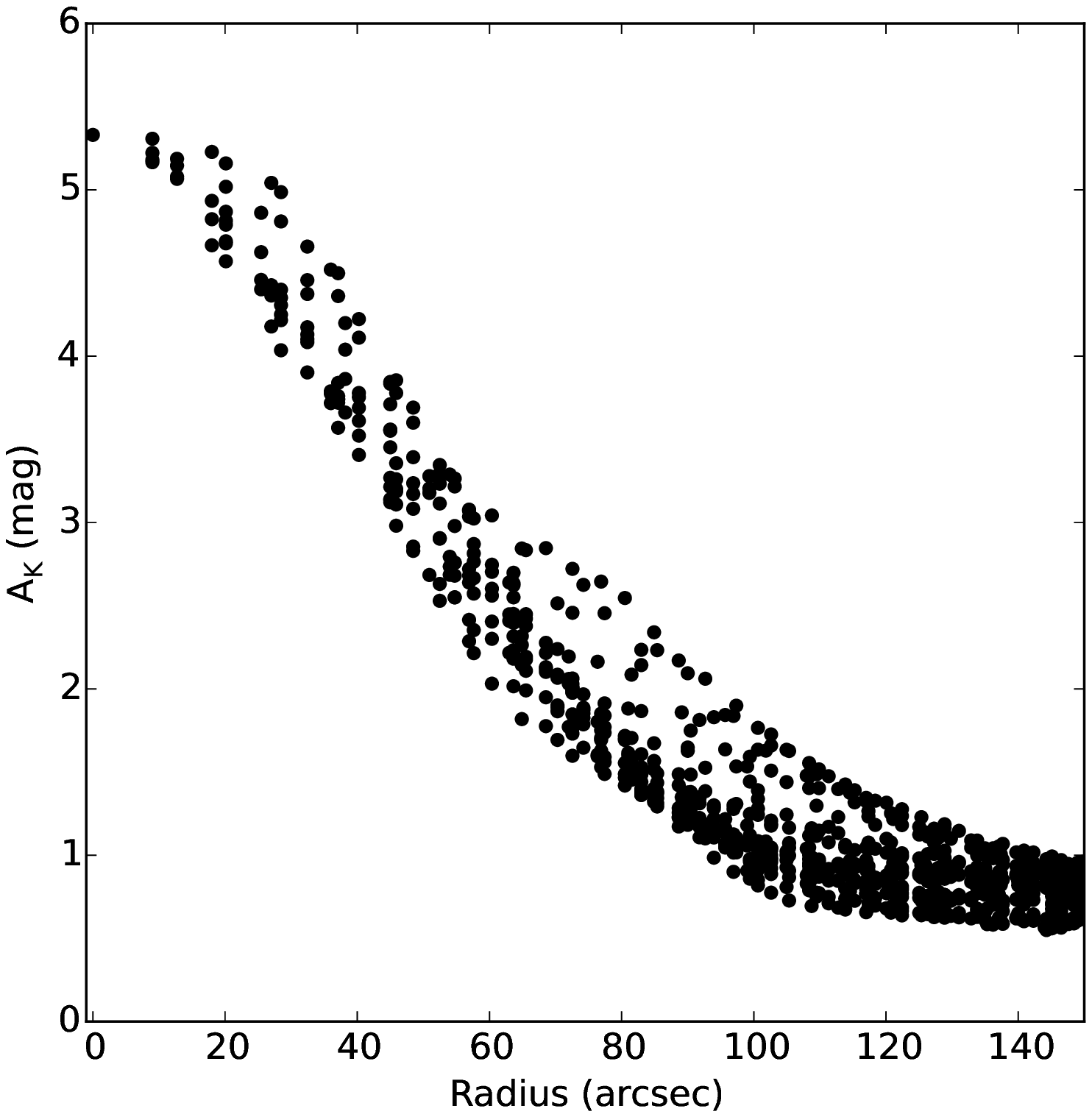}
\caption{Left: Column density profile of the FeSt 1-457 core, using an extinction map with 9$''$ resolution and $4\farcs5$ pixels. Right: Column density profile of the FeSt 1-457 core, using an extinction map with 18$''$ resolution and 9$''$ pixels.\label{fig_profile}}
\end{figure*}

Since {\it Herschel}-based column density maps require a fit to the SED, their angular resolution is limited by the band with the largest beam size. This is the SPIRE 500~$\mu$m band with a beam size of 36$''$ FWHM. However, when mapping the dust extinction instead of the dust emission, an even better angular resolution can be reached. In our VLT near-infrared data of the FeSt 1-457 core, there are enough background stars to produce a NICEST extinction map with a resolution of 9$''$ FWHM (with a minimum of about ten background stars per resolution element, see Section~\ref{sec_nir}). Fitting a 2D Gaussian to the core as a zero-order approximation confirms the slight asymmetry of the core where the larger dimension is 22\% larger than the smaller one: the 2D Gaussian has widths of $1\farcm86\times1\farcm52$ FWHM. To show the radial structure of the FeSt 1-457 core, we additionally plot the column density as a function of distance to the peak extinction pixel. Figure~\ref{fig_profile} shows the resulting profile when based on the infrared extinction profile. No annular averaging has been applied to highlight the slight ellipticity of the core. For comparison, we additionally show a profile from an extinction map with a resolution of 18$''$. While the profiles look very similar, the slight non-circular structure only becomes clearly apparent in the high-resolution extinction map. These high-resolution extinction data of the FeSt and other cores will be discussed in more detail by Rom\'an-Z\'u\~niga et al., ({\it in prep.}).

\subsection{Comparison with Barnard 68\label{sec_b68}}

We briefly test the robustness of our conclusions by applying the same methodology to a second starless core, Barnard 68, which is also located in the Pipe Nebula \citep{alv01}. Two recent studies use {\it Herschel} submillimeter imaging and subsequent modeling to construct physical models of this core, deriving an absolute central dust temperature of $8.2^{+2.1}_{-0.7}$~K or $9.3\pm0.5$~K \citep{nie12,roy14}. 

To compare Barnard~68 with FeSt~1-457, we make use of effective dust temperature and dust optical depth maps that are based on the same data as described above. These maps indicate a temperature gradient with a minimum effective dust temperature at the center of Barnard~68 of 13.07$\pm$0.09~K, slightly lower than in the case of FeSt~1-457. The {\it Planck}-derived dust opacity spectral index is $\beta_{\rm Planck}(B68)=1.71\pm0.06$.

Additionally, we reduced APEX-{\it Laboca} science verification observations of Barnard~68 from project 078.F-9012 in the same iterative procedure that we described above for the FeSt~1-457 data. These observations were obtained on July 17, 2007. For the comparison, the {\it Laboca} data were again convolved to a resolution of 36$''$ (FWHM). The rms noise in the image is at 14~mJy/beam, similar to the case of FeSt 1-457. Finally, we have constructed a near-infrared extinction map from data obtained as part of the same programs that also covered the FeSt~1-457 core.

A comparison of $\tau_{353{\rm GHz}}$ and near-infrared $A_K$ for B\,68 reveals a linear relation up to the peak value of $A_K\sim3$~mag (not shown), with larger scatter due to noise than in the case of FeSt~1-457. There is thus no evidence for a change in $\gamma$ in B\,68, but even in the FeSt 1-457 core the change is only becoming apparent at the highest levels of extinction, significantly exceeding the column densities in B\,68.

When comparing the {\it Laboca} flux density to the model prediction, we again find a linear function where $a=1.05\pm0.01$ and a zeropoint offset of $-0.130\pm0.004$~Jy/beam. When compared to FeSt 1-457, the zeropoint offset is smaller for this more isolated core, and the slope $a$ is closer to a value of one. The dust opacity spectal index $\beta$ thus is even closer to the {\it Planck} value than in the case of FeSt~1-457, with $\beta_{\rm N}{\rm(B68)} = 1.71-0.05 = 1.66\pm$0.07. As a result, the 1$\sigma$ confidence intervals of the dust opacity spectral indices derived for the FeSt~1-457 and B\,68 cores overlap.

\section{Summary and conclusions\label{sec_sum}}

The main result of this paper is our finding of evidence for grain growth in the inner regions of the FeSt~1-457 core from a comparison of {\it Herschel}-derived dust optical depths with corresponding near-infrared extinction measurements. We conclude that $1/\gamma$, i.e., the ratio of the submillimeter dust opacity and the near-infrared extinction coefficient, rises nonlinearly with extinction above about $A_K=3$~mag. The peak extinction corresponds to an estimated average volume density of $2.1\times10^5$~cm$^{-3}$. In the light of previous work \citep{asc13}, it seems likely that this change is mostly due to an increase of the submillimeter dust opacity $\kappa_\nu$ rather than a decrease of the near-infrared extinction coefficient. No comparable effect was found toward B\,68, but this core barely reaches these column densities.

Our second main finding in this paper is that toward the two dense cores FeSt 1-457 and Barnard~68, there is no evidence for variations in the dust opacity spectral index $\beta$ over a wide range of physical scales. A single value of $\beta$ is sufficient to explain the dust emission {\it within} both FeSt 1-457 and Barnard~68. There is no evidence for a varying dust opacity spectral index in either core. We find $\beta=1.53\pm0.07$ for FeSt 1-457 and $\beta=1.66\pm0.07$ for Barnard~68. The two cores thus have very similar dust opacity spectral indices. These values of $\beta$ are, finally, very similar to the numbers derived in the {\it Planck} all-sky dust model in the same locations. 

Our experiment thus gives similar results for two different cores (FeSt 1-457 and B\,68). Given the number of parameters involved, it would already be a fortuitous coincidence to produce the linearity in Figure~\ref{fig_sobs_spred} with variations in multiple parameters that would have to cancel out in their effect on $\beta$. Obtaining the same result in a second core underlines the conclusion that the absence of $\beta$ variations on the scales probed is the most simple explanation, involving just a single parameter.

The dust opacity spectral index toward the FeSt 1-457 and Barnard~68 cores thus is essentially the same whether derived on scales of $\sim$1~pc (30$'$ at a distance of 130~pc) or scales of $\sim$0.02~pc (the {\it Herschel} resolution of 36$''$). Using the easily available {\it Planck}-derived dust opacity spectral indices would constitute excellent starting points for a study of these cores. This in notable contrast to {\it Planck}-derived effective dust temperatures where the resolution does make a difference and the temperature profile of nearby cores is resolved better using {\it Herschel}-based maps. However, the absence of evidence for $\beta$ variations means that reliable effective dust temperature maps can be produced. 

Additional results and conclusions of this paper can be summarized as follows:

\begin{itemize}
\item The effective dust temperature toward the center of FeSt 1-457 is 13.5~K which is higher than the ammonia kinetic gas temperature of 9.5~K, measured toward the same core. This could be due to insufficient thermodynamic coupling of dust and gas, or, more likely, the effect of averaging along the line of sight, or both.
\item A comparison of the {\it Herschel} effective dust temperature map of FeSt 1-457 with a map of N$_2$H$^+$ emission shows that this molecular emission is constrained to the coldest and densest parts of the core, confirming previous work on a sample of Pipe cores. Maps of effective dust temperature, which are easier to obtain than millimeter line maps, thus can serve as a reasonable means of identification of the coldest and densest parts of molecular clouds in spite of the line-of-sight averaging.
\item There is a zeropoint offset in the {\it Laboca} data which is entirely due to spatial filtering of large-scale emission that is unrelated to the core, as can be shown in comparison to {\it Planck} data.
\item For a detailed study of the spatial structure of nearby cores, the resolving power of near-infrared extinction mapping remains unsurpassed when seen against the Galactic bulge. Based on near-infrared observations that were obtained with the VLT, we present a NICEST extinction map with a resolution of 9$''$ FWHM, or 0.006~pc at the distance of the Pipe Nebula. This is better by a factor of four when compared to {\it Herschel} information that is based on a convolution to the resolution of the SPIRE 500~$\mu$m band.
\end{itemize}

\begin{acknowledgements}
We thank an anonymous referee for comments that helped us to clarify and significantly improve the paper, Ralf Launhardt for insightful discussions, Arnaud Belloche, Giorgio Siringo, and Axel Weiss for help with the {\it Laboca} data analysis, and Ted Bergin for providing us with the IRAM data from the Aguti et al. (2007) paper. {\it Herschel} is an ESA space observatory with science instruments provided by European-led Principal Investigator consortia and with important participation from NASA. APEX is a collaboration between Max Planck Institut f\"ur Radioastronomie (MPIfR), Onsala Space Observatory (OSO), and the European Southern Observatory (ESO). This publication in A\&A is supported by the Austrian Science Fund (FWF).
\end{acknowledgements}

\bibliography{pipe} 

\end{document}